\documentclass[acmsmall, nonacm]{acmart}
\usepackage{listings, listings-rust}

\newcommand{\lblr}[1]{\textbf{(#1)}}
\begin{document}

\title{RustViz: Interactively Visualizing Ownership and Borrowing}

\author{Gongming (Gabriel) Luo}
\affiliation{%
  \institution{University of Michigan}
  \city{Ann Arbor}
  \state{Michigan}
  \country{USA}}
\email{luogm@umich.edu}

\author{Vishnu Reddy}
\affiliation{%
  \institution{University of Michigan}
  \city{Ann Arbor}
  \state{Michigan}
  \country{USA}}
\email{reddyvis@umich.edu}

\author{Marcelo Almeida}
\affiliation{%
  \institution{University of Michigan}
  \city{Ann Arbor}
  \state{Michigan}
  \country{USA}}
\email{mgba@umich.edu}

\author{Yingying Zhu}
\affiliation{%
  \institution{University of Michigan}
  \city{Ann Arbor}
  \state{Michigan}
  \country{USA}}
\email{zyy@umich.edu}

\author{Ke Du}
\affiliation{%
  \institution{University of Michigan}
  \city{Ann Arbor}
  \state{Michigan}
  \country{USA}}
\email{madoka@umich.edu}

\author{Cyrus Omar}
\affiliation{%
  \institution{University of Michigan}
  \city{Ann Arbor}
  \state{Michigan}
  \country{USA}}
\email{comar@umich.edu}

\begin{abstract}
    Rust is a systems programming language that guarantees memory safety without 
    the need for a garbage collector by statically tracking ownership and
    borrowing events. The associated rules are subtle and unique among industry
    programming languages, which can make learning Rust more challenging.
    Motivated by the challenges that Rust learners face, we are developing
    RustViz, a tool that allows teachers to generate an interactive timeline
    depicting ownership and borrowing events for each variable in a Rust code
    example. These visualizations are intended to help Rust learners develop an 
    understanding of ownership and borrowing by example. This paper introduces
    RustViz by example, shows how teachers can use it to generate
    visualizations, describes learning goals, and proposes a study designed to
    evaluate RustViz based on these learning goals.
\end{abstract}

\maketitle

\section{Introduction}
Rust is a systems programming language that enforces a set of rules for using
resources in memory with the concepts of \emph{ownership} and \emph{borrowing}.
These rules are enforced with the compiler's \emph{borrow checker}, which allows
Rust programs to be memory-safe without garbage collection. Instead, resources
are deallocated (``dropped'', in Rust parlance) when the static lifetime of the
owner ends.

Rust has been gaining traction. For five consecutive years, Rust was the ``most
loved'' programming language in Stack Overflow's Developer Survey
\cite{Overflow2020}. In industry, Rust has notable users such as Microsoft
\cite{Vengalil2019}, Amazon \cite{Barsky2019}, Mozilla \cite{Herman2016},
Cloudflare \cite{Kitchen2019}, Dropbox \cite{Jayakar2020}, and Discord
\cite{Howarth2020}. With Rust adoption increasing, improvements to Rust's
tooling ecosystem would be beneficial for developers.

While the ownership and borrowing rules ensure memory safety without the
performance overhead of garbage collection, these rules also make Rust difficult
to learn and Rust code difficult to reason about, even for experienced
programmers. A survey of posts on various online forums found that people
frequently complained about the borrow checker's complexity \cite{Zeng2019}. In
one of their interviews of professional programmers, \citet{Shrestha2020} found
that the borrow checker was an ``alien concept'' and ``the biggest struggle''
for an interviewee who was learning Rust. Searches on Rust forums show that
``fighting the borrow checker'' is a common phrase.

Motivated by the difficulties that Rust learners face, we are developing
\emph{RustViz}, a tool for creating visualizations depicting ownership and
borrowing events on an interactive timeline for each variable, displayed aligned
with the Rust source code. With RustViz, there are two kinds of users:
\emph{teachers} use RustViz to generate visualizations and \emph{learners}
interact with the visualizations.

We discuss RustViz from the learner's perspective in
Sec.~\ref{sec:rustviz-by-example}. We then discuss how teachers can prepare
visualizations in Sec.~\ref{sec:generating-visualizations}. In
Sec.~\ref{sec:learning-goals-and-study-design}, we propose a study that we plan
to conduct to evaluate the effectiveness of RustViz based on how well the
visualizations help Rust learners accomplish specified learning goals. We
conclude after a discussion of future and related work in
Secs.~\ref{sec:future-work}-\ref{sec:conclusion}.

\section{RustViz by Example}
\label{sec:rustviz-by-example}
RustViz shows ownership and borrowing in Rust with a timeline of memory events
for each variable in the source code that the teacher chooses to visualize. This
timeline is displayed aligned with the corresponding source code (which itself
appears inside educational material generated by the \verb|mdbook| tool, see
Sec.~\ref{sec:generating-visualizations}). When learners hover their cursor over
parts of the timeline, Hover Messages are displayed which describe the memory
events in more detail. To avoid visual clutter, we describe the key Hover
Messages in the text by referring to labels in the figures. These examples are
also available as interactive demos at:

\begin{center}
    \urlstyle{tt}\url{https://web.eecs.umich.edu/~comar/rustviz-hatra20/}
\end{center}

\subsection{First Example: Moves, Copies and Drops}
\label{sec:first-example}
\begin{figure}[h]
    \centering
    \includegraphics[width=\linewidth]{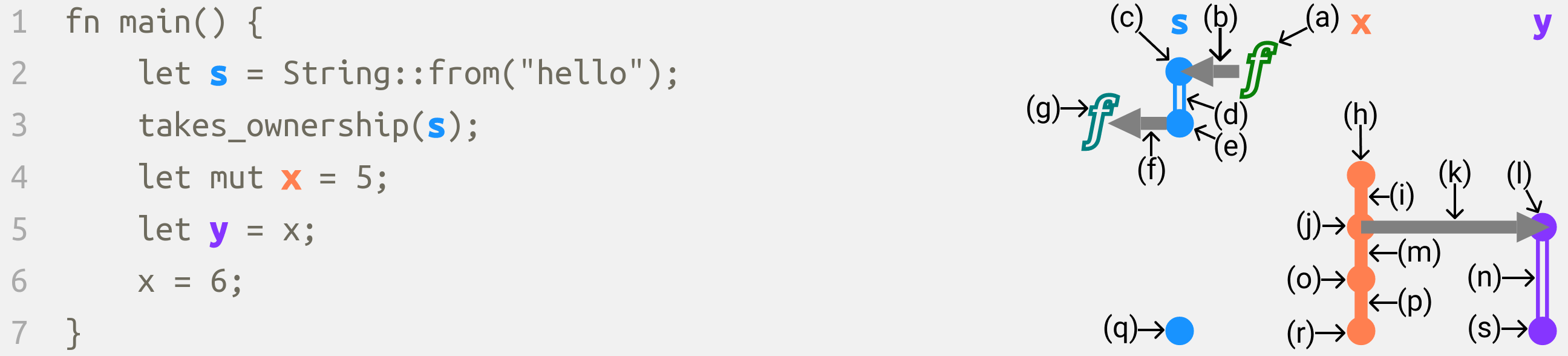}
    \vspace{-15px}
    \caption{Example visualization in RustViz with moves, copies, and drops
    (sublabels are only for exposition)}
    \Description{Example RustViz visualization demonstrating moves, copies, and
    drops (sublabels are only for exposition)}
    \label{fig:first-example}
    \vspace{-10px}
\end{figure}

Fig.~\ref{fig:first-example} shows a simple example that demonstrates 
\emph{moves}, \emph{copies}, and \emph{drops}.

\subsubsection{Moves}

On Line \textbf{2}, a \texttt{String} resource is created and bound to
\texttt{s}. In Rust, each resource has a unique \emph{owner}, and when ownership
changes, it is called a \emph{move}. In this case, \texttt{String::from}
heap-allocates and returns a \texttt{String}, which moves ownership of that
resource to \texttt{s} as indicated by the Arrow at \textbf{(b)} pointing from
\textbf{(a)} to the Dot at \textbf{(c)}. The hollow Line Segment at \textbf{(d)}
indicates that \texttt{s} cannot be reassigned nor can it be used to mutate the
resource (because \texttt{let} rather than \texttt{let mut} was used to
introduce \texttt{s}). As learners hover their mouse over these visual elements,
they see the following Hover Messages:

\begin{enumerate}
    \item[\lblr{a}] \texttt{String::from()} (and the corresponding function in
    the code is highlighted)

    \item[\lblr{b}] \textit{Move from} \texttt{String::from()} \textit{to}
    \texttt{s}
    
    \item[\lblr{c}] \texttt{s} \textit{acquires ownership of a resource}

    \item[\lblr{d}] \texttt{s} \textit{is the owner of the resource. The binding
    cannot be reassigned.}
\end{enumerate}

On Line \textbf{3}, \texttt{takes\_ownership} (not shown) is called with
\texttt{s} as a parameter, so the \texttt{String} resource gets moved from
\texttt{s} into the function. After this move, \texttt{s} is no longer valid for
use. This move is shown in RustViz with the Arrow at \textbf{(f)} from
\textbf{(e)} to \textbf{(g)}. Since \texttt{s} is no longer valid for use after
the function call, there is no Line Segment in \texttt{s}'s timeline after the
Dot. Some Hover Messages here are:

\begin{enumerate}
    \item[\lblr{e}] \texttt{s}\textit{'s resource is moved}
    
    \item[\lblr{f}] \textit{Move from} \texttt{s} \textit{to}
    \texttt{takes\_ownership()}
    
    \item[\lblr{g}] \texttt{takes\_ownership()} (and the corresponding function
    in the code is highlighted)
\end{enumerate}

\subsubsection{Copies}

On Line \textbf{4}, the (immutable) integer value 5 is bound to \texttt{x}. We
use \texttt{let mut} rather than \texttt{let}, so \texttt{x} can be reassigned.
as indicated by the solid Line Segment at \textbf{(i)}. The Hover Messages are:

\begin{enumerate}
    \item[\lblr{h}] \texttt{x} \textit{acquires ownership of a resource}
    
    \item[\lblr{i}] \texttt{x} \textit{is the owner of the resource. The binding
    can be reassigned.} 
\end{enumerate}

On Line \textbf{5}, \texttt{y} is initialized with \texttt{x}. In Rust, types
with stack-only data, like integers, generally have an annotation called the
\texttt{Copy} trait. Resources of these types get \textit{copied} rather than
moved. The copy is shown in RustViz with the Arrow at \textbf{(k)} pointing from
\textbf{(j)} to \textbf{(l)}. Since \texttt{x} is still valid for use, the solid
Line Segment continues in \texttt{x}'s timeline at \textbf{(m)}. The Line
Segment at \textbf{(n)} is hollow because we used \texttt{let} rather than
\texttt{let mut} for \texttt{y}. The Hover Messages here are:

\begin{enumerate}
    \item[\lblr{j}] \texttt{x}\textit{'s resource is copied}
    
    \item[\lblr{k}] \textit{Copy from} \texttt{x} \textit{to} \texttt{y}
    
    \item[\lblr{l}] \texttt{y} \textit{is initialized by copy from} \texttt{x}

    \item[\lblr{m}] \texttt{x} \textit{is the owner of the resource. The binding
    can be reassigned.}

    \item[\lblr{n}] \texttt{y} \textit{is the owner of the resource. The binding
    cannot be reassigned.}
\end{enumerate}

On Line \textbf{6}, we mutate \texttt{x}, as indicated by the Dot at
\textbf{(o)} continuing to solid Line Segment \textbf{(p)}:

\begin{enumerate}
    \item[\lblr{o}] \texttt{x} \textit{acquires ownership of a resource} 
    
    \item[\lblr{p}] \texttt{x} \textit{is the owner of the resource. The binding
    can be reassigned.}
\end{enumerate}

\subsubsection{Drops}

The variables \texttt{s}, \texttt{x}, and \texttt{y} go out of scope at the end of the
function on Line \textbf{7}, as indicated by Dots \textbf{(q)}-\textbf{(s)}. Since \texttt{x} and \texttt{y} were owners, their
resources are \emph{dropped}. However, since \texttt{s}'s resource was moved earlier, 
no drop occurs, as indicated in the corresponding Hover Messages:

\begin{enumerate}
    \item[\lblr{q}] \texttt{s} \textit{goes out of scope. No resource is
    dropped.}
    
    \item[\lblr{r}] \texttt{x} \textit{goes out of scope. Its resource is
    dropped.}  
    
    \item[\lblr{s}] \texttt{y} \textit{goes out of scope. Its resource is
    dropped.}  
\end{enumerate}

\subsection{Second Example: Immutable and Mutable Borrows}
\begin{figure}[h]
    \centering
    \includegraphics[width=\linewidth]{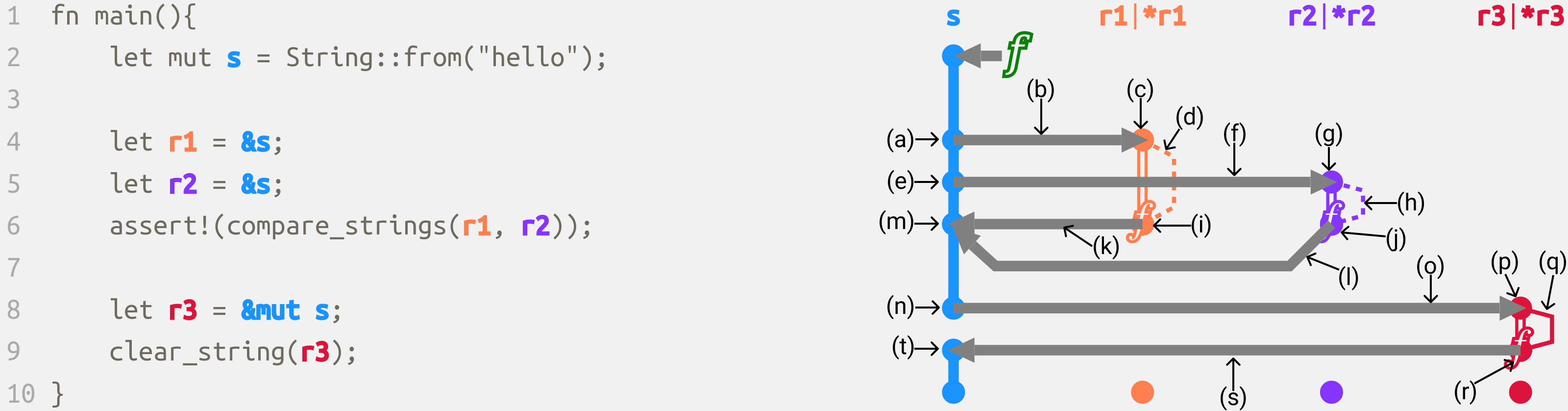}
    \vspace{-15px}
    \caption{Example visualization in RustViz with immutable and mutable
    borrows.}
    \Description{Example visualization in RustViz with immutable and mutable
    borrows.}
    \label{fig:second-example}
    \vspace{-10px}
\end{figure}

Our second example, in Fig.~\ref{fig:second-example}, demonstrates
\emph{borrowing}, i.e. working with references to resources.

\subsubsection{Immutable Borrows}

On Line \textbf{2}, a mutable \texttt{String} resource is created as described
in Sec.~\ref{sec:first-example}.

On Lines \textbf{4} and \textbf{5}, \textit{immutable references} to \texttt{s}
are created and assigned to \texttt{r1} and \texttt{r2}, respectively. In Rust,
immutable references are created with the \texttt{\&} operator rather than the
\texttt{\&mut} operator, and the creation of an immutable reference is called an
\emph{immutable borrow}. Resources cannot be mutated through immutable
references, so it is safe for Rust's borrow checker to allow multiple immutable
borrows of a resource to be live. The timelines for variables of reference type
are split into two parts. The first part is the Line Segment that represents the
variable itself, which is displayed in the same way as variables of other types.
The second part is the curved line to the right of the Line Segment, here
\textbf{(d)} and \textbf{(h)},  which represents how the borrow allows access to
the resource, i.e. how \texttt{*r1} and \texttt{*r2} work as indicated in the
header. The Hover Messages here are:

\begin{enumerate}
    \item[\lblr{a}] \texttt{s}\textit{'s resource is immutably borrowed}

    \item[\lblr{b}] \textit{Immutable borrow from} \texttt{s} \textit{to}
    \texttt{r1}

    \item[\lblr{c}] \texttt{r1} \textit{immutably borrows a resource}

    \item[\lblr{d}] \textit{Cannot mutate} \texttt{*r1}

    \item[\lblr{e}] \texttt{s}\textit{'s resource is immutably borrowed}

    \item[\lblr{f}] \textit{Immutable borrow from} \texttt{s} \textit{to}
    \texttt{r2}

    \item[\lblr{g}] \texttt{r2} \textit{immutably borrows a resource}

    \item[\lblr{h}] \textit{Cannot mutate} \texttt{*r2}
\end{enumerate}

On Line \textbf{6}, \texttt{r1} and \texttt{r2} are passed to
\texttt{compare\_strings}, which is shown in RustViz by the \textit{f} symbol on
\textbf{(i)} and \textbf{(j)}. \texttt{compare\_strings} can read the resource
through these references, but cannot mutate it. Since \texttt{r1} and
\texttt{r2}'s borrows are no longer used after the function returns, the
borrowed resource is returned, despite \texttt{r1} and \texttt{r2}'s lexical
scope not ending. This is due to a feature in Rust called \emph{non-lexical
lifetimes}, which allows more programs to pass the borrow checker by
disregarding borrows which are no longer live because they are never used again.
The return of these borrows are represented by the Arrows at \textbf{(k)} and
\textbf{(l)}. Some Hover Messages here are:

\begin{enumerate}
    \item[\lblr{i}] \texttt{compare\_strings()} \textit{reads from} \texttt{r1}

    \item[\lblr{j}] \texttt{compare\_strings()} \textit{reads from} \texttt{r2}

    \item[\lblr{k}] \textit{Return immutably borrowed resource from} \texttt{r1}
    \textit{to} \texttt{s}

    \item[\lblr{l}] \textit{Return immutably borrowed resource from} \texttt{r2}
    \textit{to} \texttt{s}

    \item[\lblr{m}] \texttt{s}\textit{'s resource is no longer immutably
    borrowed}
\end{enumerate}

\subsubsection{Mutable Borrows}

On Line \textbf{8}, a \emph{mutable reference} to \texttt{s}'s resource is
created and assigned to \texttt{r3}. In Rust, mutable references are created
with the \texttt{\&mut} operator rather than the \texttt{\&} operator, and the
creation of a mutable reference to a resource is called a \emph{mutable
borrow}. Resources can be mutated through mutable references. To ensure memory
safety, if a mutable borrow of a resource is live, no other borrow of that
resource, mutable or immutable, may be live. Thanks to non-lexical lifetimes,
\texttt{r3}'s borrow does not conflict with \texttt{r1} or \texttt{r2} because
\texttt{r1} and \texttt{r2}'s borrows are not live. We show that \texttt{r3}'s
borrow is mutable in RustViz with a Solid Line at \textbf{(q)}. To ensure memory
safety, Rust does not allow a resource to be accessed through its owner if a
mutable borrow of that resource is live. Since \texttt{r3} mutably borrows
\text{s}'s resource, we cannot access the resource through \texttt{s}, so there
is no Line Segment between Lines \textbf{6} and \textbf{7}. Some Hover Messages
here are:

\begin{enumerate}
    \item[\lblr{n}] \texttt{s}\textit{'s resource is mutably borrowed}

    \item[\lblr{o}] \textit{mutable borrow from} \texttt{s} \textit{to}
    \texttt{r3}

    \item[\lblr{p}] \texttt{r3} \textit{mutably borrows a resource}

    \item[\lblr{q}] \textit{Can mutate the resource *r3}
\end{enumerate}

On Line \textbf{9}, \texttt{r3} is passed to \texttt{clear\_string}, and
\texttt{clear\_string} is able to mutate the \texttt{String} through that
reference. After the function returns, the borrowed resource is returned to
\texttt{s}. With the \texttt{String} resource no longer being mutably borrowed,
\texttt{s}'s Line Segment in the timeline resumes. Some Hover Messages here are:

\begin{enumerate}
    \item[\lblr{r}] \texttt{clear\_string()} \textit{reads from} \texttt{r3}

    \item[\lblr{s}] \textit{Return mutably borrowed resource from} \texttt{r3}
    \textit{to} \texttt{s}

    \item[\lblr{t}] \texttt{s}\textit{'s resource is no longer mutably borrowed}
\end{enumerate}

At the end of the function on Line \textbf{10}, \texttt{s}, \texttt{r1},
\texttt{r2}, and \texttt{r3} go out of scope. Since \texttt{s} is the owner of a
resource, the resource is dropped.

\subsubsection{Ownership and Borrowing for Variables of Reference Type}
In Rust, variables of reference type follow ownership and borrowing rules that
variables of other types do. Ownership of mutable references can be moved, while
ownership of immutable references get copied, since immutable references have
the \texttt{Copy} trait. Variables of reference type can be mutably and
immutably borrowed.

In the above visualization, some of these subtleties are not shown. For example,
on Line \textbf{6}, the references of \texttt{r1} and \texttt{r2} are copied
into \texttt{compare\_strings} since immutable references have the \texttt{Copy}
trait. However, we choose not to show these copies in the visualization. On Line
\textbf{9}, the reference of \texttt{r3} is moved into \texttt{clear\_string}.
Again, we choose not to show this. 

The reason for this is to focus on showing the ownership and borrowing of the
\texttt{String} resource, which is most pertinent to a learner who is just being
introduced to the ownership and borrowing system. If we show every single memory
event, the visualization would become too noisy and potentially distract the
learner from the focus of the example. One advantage of our approach of using
manually-determined memory events rather than automatic generation from source
code is that the teacher can choose precisely which memory events to focus on.

\section{Generating Visualizations}
\label{sec:generating-visualizations}
With RustViz, teachers generate visualizations using our Rust library. The
library is still a work in progress, and there are certainly improvements that
can be made to the workflow of generating a visualization. In this section, we
walk through a simple example of how a teacher would currently use our tool to
generate the following very simple visualization:

    \includegraphics[width=0.4\linewidth]{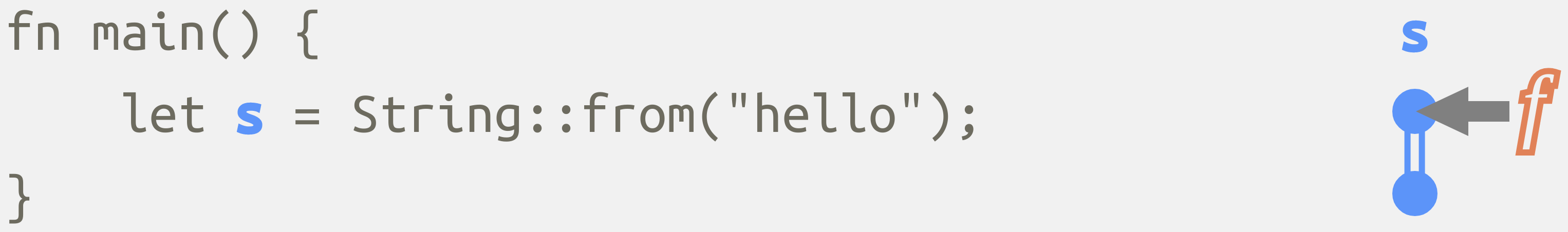}

First, the teacher creates an annotated version of the code marking variables
and functions with unique hashes (the data hash is always \texttt{0} for
functions). Each use of the same variable uses the same hash.

\begin{lstlisting}[columns=spaceflexible, basicstyle=\ttfamily\scriptsize]
    fn main() {
        let <tspan data-hash="1">s</tspan> =
        <tspan class="fn" data-hash="0" hash="2">String::from</tspan>("hello");
    }
\end{lstlisting}

Once the annotated version of the source code is created, teachers write a
visualization specification that calls RustViz library functions. In this case,
the code to generate the visualization is as follows:

\begin{lstlisting}[language=Rust, style=colouredRust, basicstyle=\ttfamily\scriptsize]
    //variable s
    let s = ResourceAccessPoint::Owner(Owner {
        hash: 1, name: String::from("s"), is_mut: false, lifetime_trait: LifetimeTrait::None });
    //function String::from()
    let from_func = ResourceAccessPoint::Function(Function {
        hash: 2, name: String::from("String::from()") });
    //data structure that will contain our specified memory events
    let mut vd = VisualizationData {
        timelines: BTreeMap::new(), external_events: Vec::new() };
    //specify the memory events
    vd.append_external_event(ExternalEvent::Move{
      from: Some(from_func), to: Some(s.clone())}, &(2 as usize));
    vd.append_external_event(ExternalEvent::GoOutOfScope{ ro: s.clone() }, &(3 as usize));
    //render the visualization
    svg_generation::render_svg(&"04_01_01".to_owned(), &"one_var".to_owned(), &vd);
\end{lstlisting}

We first describe the variable \texttt{s}. The hash and name matches what was
seen in the annotated source code. We set \texttt{is\_mut} as \texttt{false}
because \texttt{let} was used rather than \texttt{let mut} in the source code.
We set \texttt{lifetime\_trait} as \texttt{LifetimeTrait::None} because there
are no moves or copies from \texttt{s} that we want to show in the
visualization. If we wanted to show any moves from \texttt{s}, we would set this
to \texttt{LifetimeTrait::Move}. If we wanted to show any copies from
\texttt{s}, we would set this to \texttt{LifetimeTrait::Copy}. We then describe
the function \texttt{String::from()}. For functions, we only need to specify the
hash and name. 

After the variables are described, we create the data structure that will hold
the specified memory events. The first event is a move from
\texttt{String::from()} to \texttt{s}, which we specify by giving where the move
is from, where the move is to, and on what line the move occurs. The second
event is \texttt{s} going out of scope. 

Finally, we call \texttt{render\_svg} which renders the visualization as an SVG.
The first two arguments tell the library where to find the annotated source code
and the last argument is the data structure that we added our specified memory
events to. This SVG can be included in a book generated using the
\texttt{mdbook} system, which is used for generating the Rust Book
\cite{Klabnik2018} and other Rust documentation. The SVG can be tweaked after it
is generated in situations where the teacher wants more fine-grained control
over the visualization.

Although somewhat complex, it is reasonably straightforward to create new 
visualizations given examples of existing visualizations. The benefit of manual
construction is that the teacher can omit events that may not be relevant to the
learning goal of the example.

\section{Learning Goals and Proposed Study Design}
\label{sec:learning-goals-and-study-design}

\subsection{Learning Goals}

To evaluate the effectiveness of RustViz as a learning tool, we plan to conduct
human studies to see if RustViz measurably helps learners develop a mental model
for reasoning about Rust code. To that end, we determined a set of learning
goals that RustViz is intended to help learners accomplish and will be used as
the evaluation criteria: 
\begin{enumerate}
    \item Understand concepts related to ownership in Rust
    \begin{enumerate}
        \item Understand that each resource has a unique identifier called its
            owner
        \item Identify when resources are dropped
        \item Understand the different ways that ownership can be moved
        \item Understand the difference between move and copy
    \end{enumerate}
    \item Understand Rust's borrowing rules
    \begin{enumerate}
        \item Understand the difference between mutable and immutable borrows
        \item Understand how ownership and borrowing interact with each other
    \end{enumerate}
    \item Understand the difference between lexical scope and non-lexical
        lifetimes
\end{enumerate}

Notably, we avoid certain concepts such as lifetime annotations, compound data
types (e.g., structs and array slices), reborrowing, and the \texttt{unsafe}
keyword. The focus of RustViz is to provide learners a good introduction to the
\emph{basics} of Rust's ownership and borrowing system. 

\subsection{Study Format}

The participants in our study will be adults who are proficient in C or C++ but
have no prior exposure to Rust, screened out through self-reporting. During the
study, participants will be asked to complete three tasks: the pre-survey, the
interactive tutorial, and the post-survey.

In the pre-survey, participants will answer questions about their experience and
familiarity with various programming languages and programming concepts.

In the interactive tutorial, participants will view a series of short readings
that introduce basic Rust concepts. Following each reading, participants will
answer questions related to the reading's topic, with the ability to refer back
to the reading if desired. Questions may draw on concepts from a prior section. 
We will impose a time limit for each section that participants should be able to
comfortably stay within. To measure the effect of RustViz, some participants
will view readings without visualizations (the control group), while others will
view readings with visualizations accompanying any example code (the treatment
group). The readings will provide sufficient information to answer the questions
even without visualizations.

After the interactive tutorial, participants will be prompted to complete a
post-survey asking about their perceptions regarding the tool.

Due to the time constraints of a study, this setting is likely not the most
conducive to learning the material. However, we would still like to conduct the
study to see if there are measurable differences between those who view the
visualizations and those who do not. The qualitative feedback we receive from
the study may also be helpful in finding possible improvements to RustViz.
Later, we plan to conduct similar studies in a longer-term classroom setting.

\subsection{Interactive Tutorial}

As the participants go through the interactive tutorial, we will record the time
it takes for them to complete each section as well as the accuracy of their
responses to the questions. By classifying questions and sections by the Rust
concepts they cover (e.g., mutability, borrowing, ownership) we will be able to
evaluate whether the visualization helps participants achieve the identified
learning goals. In particular, we believe that decreased completion time and
increased accuracy in the treatment group for questions and sections related to
a specific learning goal is evidence that RustViz helps learners achieve that
learning goal. One example of a question we might ask in the interactive
tutorial is as follows:

\noindent
\textit{Consider the following Rust program.}
\begin{lstlisting}[language=Rust, style=colouredRust, basicstyle=\ttfamily\scriptsize]
    fn id(y : String) -> String {y}
    fn f(x : String) -> i32 {
      println!("Hello, {}!", x);
      let z = id(x);
      println!("Goodbye, {}!", z);
      42
    }
    fn main() {
      println!("Welcome!");
      {
        let a = String::from("world");
        println!("Thinking...");
        let q = f(a);
        println!("The meaning of life is: {}.", q);
      }
      println!("Done.");
    }
\end{lstlisting}

We would then provide the participant with the text sent to standard out and ask
``\textit{The string resource that} \texttt{a} \textit{initially owns is dropped
between which two adjacent lines of output above?}''.

As a follow up, we could ask the participant ``\textit{Describe how ownership of
the string resource is moved. Your answer should take the form of a sequence of
events chosen from the drop-down list. You do not need to account for calls to}
\texttt{println!} \textit{(it is a macro and does not cause a move).}''

\subsection{Other Collected Data}
We will also record the cursor movements of the participants, which will allow
us to measure usage of the interactive features in RustViz. This data will be
used to evaluate whether increased usage of the interactive features helps
participants achieve the identified learning goals. 

We will use the post-survey to get qualitative feedback on RustViz. One example
of a question would be to ask them to rate their agreement with the statement
\textit{``The visualizations helped me develop an understanding of Rust's
borrowing rules''}, which would help us further evaluate RustViz against
learning goal 2. We will also use the post-survey to collect feedback on both
the visualization and the study design, so we can improve them in the future. 

\section{Future Work}
\label{sec:future-work}
The next step would be to conduct a pilot study as described in
Sec.~\ref{sec:learning-goals-and-study-design} to better understand the
effectiveness of RustViz as a learning tool. If the study reveals possible
improvements to RustViz, we may implement them and conduct further studies. We
also plan to use RustViz in an undergraduate programming languages course during
a unit on Rust. This will provide us with more data and feedback on the
effectiveness our tool.

With our current design, we did not put thought toward visualizing code that
fails the borrow checker or has branch points. It may be useful to generate a
visualization that shows violations of ownership and borrowing rules, as example
erroneous code can be useful to learners. Furthermore, it may be useful to for
learners to see how ownership and borrowing rules manifest in code with branch
points. In future revisions, we may improve our design to show these situations. 

Currently, RustViz generates visualizations from manually-determined memory
events rather than from the source code directly. We could explore the
possibility of automatic generation from source code.  However, there will be
challenges with scaling up the visualization for complex code with more
variables, branches, moves, and borrows. Moreover, the teacher may not want to
visualize everything. For these reasons, it may not be worth pursuing fully
automatic generation. Instead, creating a simplified domain-specific language
for manually generating the visualizations may be more helpful for teachers.
Automatic generation would be most useful for practitioners who want to generate
visualizations for their own code.

\section{Related Work}
\label{sec:related-work}
We find that there is little peer-reviewed research on visualizations to assist
in reasoning about Rust's unconventional rules despite interest in the Rust
community for such tools. There do exist blog posts that mock up ideas for
visualizations of Rust code. One post showcases a vertical timeline of ownership
and borrowing events \cite{Ruffwind2017}---this design is visually similar to
RustViz, except without interactivity. This blog post simply provides a mock-up
of an idea, while our contribution includes a tool for creating visualizations.
A different mock-up in another blog post shows a possible approach for
visualizing Rust code in an editor rather than for documentation
\cite{Walker2019}.

On the Rust Internals Forum, \citet{Faria2019} (username Nashenas88) started a
thread to discuss ideas for visualizing ownership and borrowing within an
editor. The thread contains various ideas for such a visualization from Faria
and others. Notably, Faria was able to create a working prototype of lifetime
visualization in Atom that is generated directly from Rust source code. This
approach works automatically from source code for use in an editor, while our 
tool works from manually-determined memory events for use in documentation.
Furthermore, this prototype shows the information by highlighting or outlining
relevant source code, rather than using a timeline approach. The prototype is no
longer being maintained, and it is unclear what the direction of the project is.

For their bachelor thesis, \citet{Dominik2018} created an algorithm that
identifies lifetime constraints and the code that generates the constraints from
information extracted from the Rust compiler, including the Polonius borrow
checker. Dominik also shows a visualization of the lifetime constraints as a
directed graph, though it is quite complex.
\citet{Blaser2019} builds on this work for their bachelor thesis by developing a
tool that can explain lifetime errors in code given the lifetime constraints.
Blaser also created an extension for Visual Studio Code that can show a
graph-based visualization of the lifetime constraints that is simpler Dominik's.
As an alternative to the graph-based visualization, Blaser's Visual Studio Code
extension can show a text-based explanation of lifetime errors in Rust code. The
focus of Blaser's tool seems to be to help programmers reason about
lifetime-related errors in their code through editor integrations, while our
focus is on generating visualizations in documentation for use in a teaching
setting. 

Outside of Rust, there are many visualization systems developed to teach
beginners about program behavior, and \citet{Sorva13} published an extensive
review of such systems. There are also tutoring systems for teaching
programming, some of which include visualizations. \citet{Crow18} created a
review of a number of these systems and identified those that have
visualizations.

\section{Conclusion}
\label{sec:conclusion}
In this paper, we introduced RustViz, the tool we are developing to generate
visualizations of ownership and borrowing in Rust programs from
manually-provided memory events. These visualizations are designed to be
displayed alongside example code in documentation to help Rust learners develop
a basic understanding of Rust's ownership and borrowing rules. We showcased our
current visual design by walking through a couple of examples, showed how
RustViz is used to generate the visualizations, described learning goals that 
RustViz is being designed to help learners accomplish, and described the design
of the pilot study we plan to conduct as an initial evaluation of our tool. Our
belief is that RustViz can improve the learning outcomes of Rust learners, and
the next step is to conduct a pilot study to better understand the effectiveness
of the tool.

\bibliographystyle{ACM-Reference-Format}
\bibliography{references}

\end{document}